\documentclass[10pt,conference]{IEEEtran}
\IEEEoverridecommandlockouts
\usepackage{color,soul}
\usepackage{cite}
\usepackage{amsmath,amssymb,amsfonts}
\usepackage{algorithmic}
\usepackage{graphicx}

\usepackage{textcomp} 
\usepackage{amsmath}
\usepackage{booktabs} 
\usepackage{multirow}
\usepackage{tikz}

\usepackage{orcidlink}
\usepackage{comment}
\usepackage{amssymb}
\usepackage{braket}
\usepackage{graphicx}
\usepackage{booktabs}
\usepackage{multirow}
\usepackage{adjustbox}
\usepackage{enumitem}
\usepackage[ruled,vlined,linesnumbered]{algorithm2e}
\usepackage{xcolor}
\def\BibTeX{{\rm B\kern-.05em{\sc i\kern-.025em b}\kern-.08em
    T\kern-.1667em\lower.7ex\hbox{E}\kern-.125emX}}

\usepackage[compact]{titlesec}

\begin{document}

\title{LEP-QNN: Loan Eligibility Prediction using Quantum Neural Networks}

\author{\IEEEauthorblockN{Nouhaila Innan\textsuperscript{1,2}, Alberto Marchisio\textsuperscript{1,2}, Mohamed Bennai\textsuperscript{3}, and Muhammad Shafique\textsuperscript{1,2}
 }
 \IEEEauthorblockA{\textsuperscript{1}eBRAIN Lab, Division of Engineering, New York University Abu Dhabi (NYUAD), Abu Dhabi, UAE\\ \textsuperscript{2}Center for Quantum and Topological Systems (CQTS), NYUAD Research Institute, NYUAD, Abu Dhabi, UAE\\ 
\textsuperscript{3}Quantum Physics and Spintronics Team, LPMC, Faculty of Sciences Ben M'sick,\\ Hassan II University of Casablanca, Morocco\\ nouhaila.innan@nyu.edu, alberto.marchisio@nyu.edu, mohamed.bennai@univh2c.ma, muhammad.shafique@nyu.edu\\ }}

\maketitle

\begin{abstract}
\textbf{Predicting loan eligibility with high accuracy remains a significant challenge in the finance sector. Accurate predictions enable financial institutions to make informed decisions, mitigate risks, and effectively adapt services to meet customer needs. However, the complexity and the high-dimensional nature of financial data have always posed significant challenges to achieving this level of precision. To overcome these issues, we propose a novel approach that employs Quantum Machine Learning (QML) for Loan Eligibility Prediction using Quantum Neural Networks (LEP-QNN).
Our innovative approach achieves an accuracy of 98\% in predicting loan eligibility from a single, comprehensive dataset. This performance boost is attributed to the strategic implementation of a dropout mechanism within the quantum circuit, aimed at minimizing overfitting and thereby improving the model's predictive reliability. In addition, our exploration of various optimizers leads to identifying the most efficient setup for our LEP-QNN framework, optimizing its performance. We also rigorously evaluate the resilience of LEP-QNN under different quantum noise scenarios, ensuring its robustness and dependability for quantum computing environments. This research showcases the potential of QML in financial predictions and establishes a foundational guide for advancing QML technologies, marking a step towards developing advanced, quantum-driven financial decision-making tools.}

\end{abstract}
\section{\label{sec:level1}Introduction}
In finance, the ability to accurately predict loan eligibility is a pivotal factor driving the dynamics of lending and borrowing, affecting financial institutions and the broader economic landscape. Assessing loan eligibility is a complex process involving analyzing vast arrays of data to understand and predict an applicant's financial behavior and repayment capacity. Although effective to a certain extent, conventional models often struggle with financial datasets' nonlinear and high-dimensional nature. Furthermore, as the financial industry evolves, driven by innovations in fintech and a growing emphasis on financial inclusion, the need for more sophisticated, accurate, and efficient predictive models becomes increasingly apparent.

Machine Learning (ML) technologies have conducted in a new era in predictive analytics, furnishing tools with the capability to learn from data and make informed predictions. These models find applications across a spectrum of financial operations, from credit scoring and fraud detection to the nuanced task of predicting loan eligibility \cite{west2000neural,ngai2011application,murugan2023large,kumar2022customer}. However, despite their considerable successes, these conventional ML paradigms are limited by the intrinsic computational confines of classical computing, especially when confronted with the complexity of large-scale financial datasets \cite{byrapu2023big,lwakatare2020large}. This computational impasse, exacerbated by the burgeoning volume and diversity of financial data, calls for a leap into computational frameworks that stride beyond the limitations of conventional computing methods.

Quantum Machine Learning (QML) stands at this frontier as a groundbreaking integration of Quantum Computing (QC) with ML algorithms \cite{biamonte2017quantum,schuld2015introduction}.
This innovative integration uses the principles of quantum mechanics to enhance ML models' processing power and efficiency \cite{innan2024variational,innan2023enhancing,innan2024fedqnn}, opening up possibilities across various domains, from materials science to healthcare 
\cite{innan2024quantum,innan2023quantum,dutta2024aq,pathak2024resource,innan2025qnn,innan2025optimizing}. Notably, its application within the finance sector has demonstrated profound impacts \cite{innan2023financial,innan2024financial,dutta2024qadqn}, underscoring the versatility and efficacy of QML in addressing complex, data-intensive challenges.
 
This insight is especially relevant in loan eligibility prediction, where discerning the subtle interplays and correlations within extensive datasets can amplify predictions' accuracy.
In the finance sector, QML models can explore the probabilistic fabric of financial markets and consumer behavior with unprecedented depth and precision. 
This insight is especially relevant in loan eligibility prediction, where discerning the subtle interplays and correlations within extensive datasets can amplify predictions' accuracy. However, QML models generally suffer from overfitting, especially when dealing with complex data \cite{schetakis2024quantum}. The application of QML to loan eligibility prediction remains largely unexplored, presenting a significant opportunity to develop new methods that provide deep insights and address the challenge of overfitting to improve financial decision-making processes.
\begin{figure}[htpb]
    \centering
    \includegraphics[width=1\linewidth]{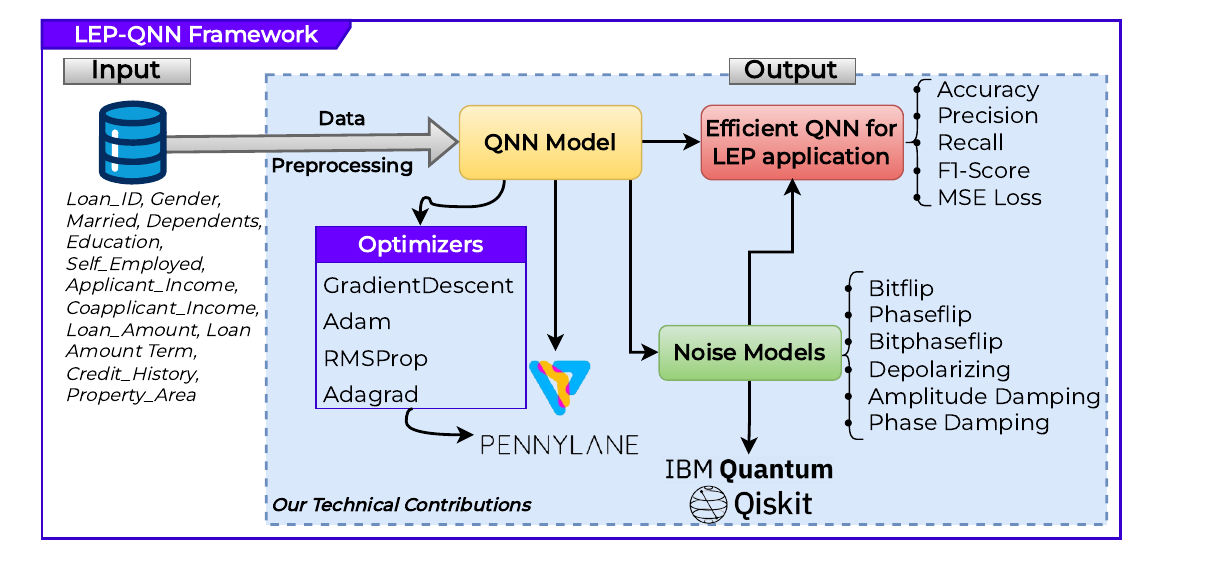}
    \caption{Overview of the LEP-QNN framework, the diagram encompasses the QNN model's interaction with various noise models, optimized by a selection of advanced optimizers within the IBM Qiskit and Pennylane platforms. This encapsulation underscores our novel contributions to QML for loan eligibility prediction.}
    \label{fig1}
\end{figure}
This paper discusses Loan Eligibility Prediction using Quantum Neural Networks (LEP-QNN), a trailblazing QML framework engineered to navigate the complexities of loan eligibility prediction. \textbf{Our contributions, illustrated in Fig.~\ref{fig1}, are summarized as follows:}
\begin{itemize}[leftmargin=*]
\item \textbf{A dropout} mechanism is seamlessly integrated into the LEP-QNN framework, aimed at reducing overfitting and boosting the framework's ability to generalize across diverse datasets. 
\item Through comprehensive testing with various \textbf{optimizers}, we identify the most effective strategies for optimizing our LEP-QNN.
\item We investigate the impact of various \textbf{quantum noise models} on the LEP-QNN's performance, offering insights into the resilience of quantum models in noisy environments. 
\end{itemize}
Our LEP-QNN framework embodies a significant advancement toward achieving more accurate, efficient, and inclusive financial services. By forging a nexus between QC and financial analytics, our study makes a notable contribution to the theoretical landscape of QML and underscores its practical implications in revolutionizing the financial industry.
This paper is organized as follows. In Sec.~\ref{sec2}, we review background and related work on QC in finance.
In Sec.~\ref{sec3}, we introduce the LEP-QNN framework and its components. In Sec.~\ref{sec4}, we present results and discussions on optimizer performance and noise model analysis. Finally, in Sec.~\ref{sec5}, we conclude with a summary of findings and future directions.

\section{Background and Related Work \label{sec2}} 
\subsection{QC and QML in Finance}
Against the backdrop of rapid technological advancements, the finance industry stands at the threshold of a revolutionary transformation, primarily driven by the emergence of QC. This paradigm shift is powered by the intrinsic capabilities of quantum computers to process information at unprecedented speeds, promising to address some of the most complex problems in finance with remarkable efficiency. Among these, QML emerges as a pivotal technology to forge a new path in financial analysis, optimization, and predictive modeling.
The potential applications of QC in finance are vast and varied \cite{breeden2023classical,solikhun2023analysis}, with state-of-the-art developments, substantial work has been focused on using quantum speedups for Monte Carlo methods, portfolio optimization, and ML. 

These advancements were epitomized by recent explorations undertaken by teams such as QC Ware, which delineated the quantum speedups achievable across these applications \cite{bouland2020prospects}.
Parallel to these developments, quantum-inspired evolutionary algorithms (QIEA) demonstrated their distinction in feature selection for credit scoring \cite{chen2024quantum}. 
Furthermore, applying optimization algorithms and quantum annealers in financial decision-making opened new avenues for efficiency gains \cite{orus2019quantum}. 
Quantum-enhanced ML models, particularly in financial risk management, have shown competitive performance against classical benchmarks \cite{leclerc2023financial}. Notably, these models offer improved interpretability and training efficiency, providing a compelling case for integrating QC with ML algorithms to develop more effective financial models.

A detailed exploration of quantum algorithms for finance revealed promising avenues for simulation, optimization, and ML applications \cite{egger2020quantum}. Practical demonstrations on platforms such as IBM Quantum back-ends provided valuable insights into the potential benefits and technical challenges of adopting QC in financial services, pointing to a need for continued development and scalability.

The practical application of QML for solving real-world financial problems was further highlighted by resources offering guidance on exploiting quantum algorithms within a Python environment \cite{saxena2023financial}. This approach underscored the advantages of QC in financial analysis and modeling, advocating for an integrated approach to harnessing the full power of quantum and classical algorithms.
In credit scoring and classification, the efficiency of feature selection was notably enhanced by applying Quadratic Unconstrained Binary Optimization (QUBO) models on quantum annealers \cite{milne2017optimal}. 

The exploration of fairness in QNNs introduced a balanced approach to deploying fair and accurate QNNs in finance \cite{wang2024justq}. This research highlighted the importance of incorporating fairness into the accuracy-driven domain of QC, proposing comprehensive solutions that balance both objectives.
Finally, developing a QC method for calculating the value at risk for portfolios demonstrated the method's effectiveness \cite{de2023var}, showcasing the practical applicability of quantum technologies in accurately managing large asset portfolios.

\subsection{Loan Eligibility Prediction}
\begin{figure*}[htpb]
    \centering
    \includegraphics[width=0.9\linewidth]{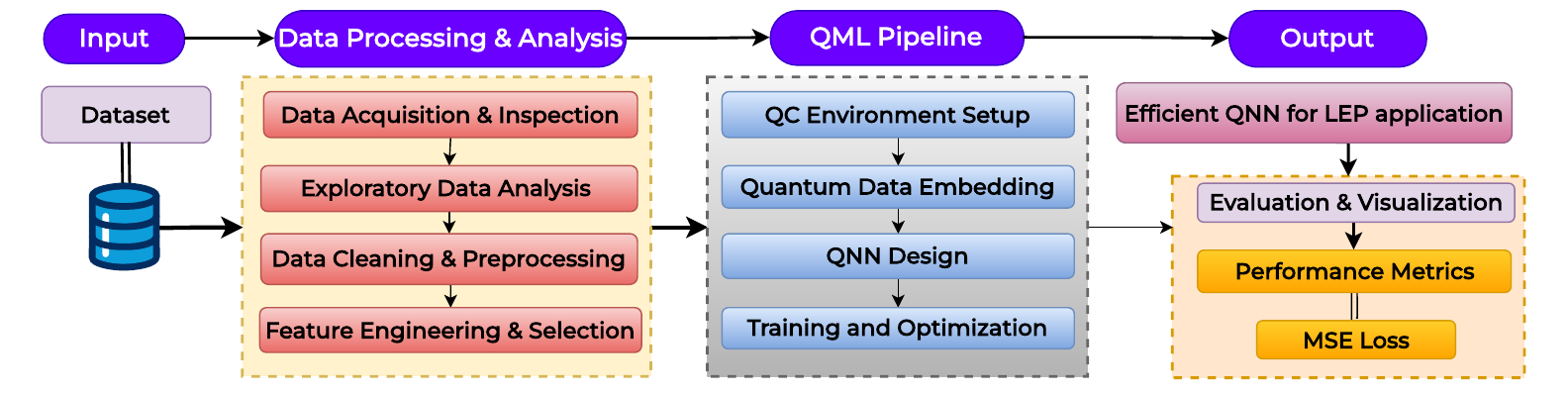}
    \caption{The LEP-QNN framework operational flowchart, outlining the integrated pipeline from dataset ingestion to predictive output. The initial phase encompasses a rigorous data treatment process. Central to the pipeline is the QC environment setup, where the nuances of quantum computation are discussed. The QNN Design phase is iterative, incorporating feedback from subsequent training and optimization to refine the model. The output stage is characterized by a dual focus on accuracy and interpretability, with performance metrics providing insights into the model's efficacy and MSE loss, offering a quantitative measure of prediction precision.}
    \label{fig2}
\end{figure*}

Collectively, these developments underscore the transformative potential of QC in finance. As QML continues to evolve, it promises to redefine the landscape of financial services, offering more accurate, efficient, and inclusive solutions. Amidst the broad spectrum of applications where QML has demonstrated its efficacy, one area remains conspicuously unexplored: the direct application of QML for loan eligibility predictions. This gap in the literature and practice highlights an opportunity for groundbreaking research and development.
The exploration of loan eligibility predictions using QML represents a novel frontier in financial analytics. Despite QML's proven capabilities in various sectors within finance, such as portfolio optimization \cite{mugel2022dynamic}, fraud detection \cite{innan2024qfnn}, and risk management \cite{leclerc2023financial}, its application for predicting loan eligibility has not been directly addressed. This oversight presents a unique challenge and opportunity for our study. By developing our LEP-QNN framework, we aim to fill this gap by enhancing predictive accuracy and computational efficiency in loan eligibility assessments.

The relevance of this endeavor is illuminated by the extensive work already conducted in the domain of ML for loan eligibility. For instance, utilizing XGBoost and random forest algorithms showcased high efficacy in classifying loan applicants based on risk, setting a precedent for advanced predictive models in this area \cite{alagic2024machine}. Furthermore, innovative clustering-based classification algorithms demonstrated potential for superior classification performance in mixed data scenarios relevant to loan eligibility \cite{kuo2024ensemble}.

The critical issue of loan default prediction has also been addressed through various ML methods, including back propagation neural networks and K Nearest Neighbor (KNN) algorithms, revealing the importance of early detection of potential default risks \cite{gao2024financial}. Similarly, research projects utilizing logistic regression, KNN, and decision trees have further confirmed the viability of ML techniques in predicting loan defaults among applicants \cite{ali2024banking}.

A comprehensive study highlighted the effectiveness of extra trees and ensemble voting models in predicting bank loan defaulters, offering insights into enhancing bank loan approval processes \cite{uddin2023ensemble}. Additionally, the comparative analysis of multiple logistic regression, decision trees, random forests, and other ensemble methods for loan default prediction identified the significant impact of accurate models on financial organizations \cite{anand2022prediction}.

In the context of growing economies, deploying ML models to determine safe loan allocations demonstrated the potential to reduce non-performing assets for banks and Nonbank financial companies (NBFCs), with Naïve Bayes models showing promising performance in loan forecasting \cite{blessie2019exploring}.

In light of these studies, our LEP-QNN framework seeks to pioneer the integration of QC and QML in loan eligibility prediction. Building on the groundwork laid by existing approaches, the quantum-enhanced capabilities of our framework aim to surpass existing methodologies in accuracy and efficiency, marking a significant leap forward in the field of financial analytics.
\begin{figure*}[h!]
    \centering
    \includegraphics[width=1\linewidth]{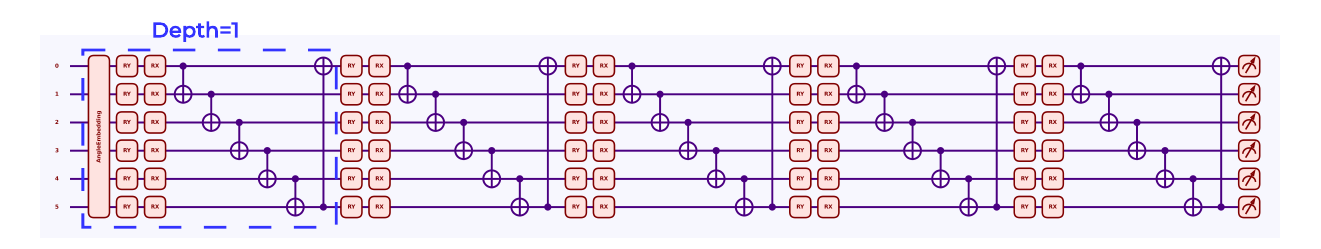}
    \caption{Visualization of the QNN circuit used in the LEP-QNN framework. The full circuit is shown at the top, with each layer comprising angle encoding followed by a series of parameterized $RY$ and $RX$ gates, and entangling $CNOT$ gates arranged in a ring pattern.} 
    \label{qnn}
\end{figure*}
\section{LEP-QNN Framework}\label{sec3}
In our methodological approach, our LEP-QNN framework capitalizes on the principles of quantum computation to significantly enhance predictive capabilities. As represented in Fig.~\ref{fig2}, the entire workflow of our framework, beginning with the initial dataset processing, through the intricate stages of the QML pipeline, and culminating in generating an output that not only forecasts loan eligibility but also offers critical performance insights. This comprehensive end-to-end process is deliberately orchestrated to ensure that every phase contributes to the ultimate objective: a model that delivers high predictive accuracy while strongly emphasizing interpretability and validation.

\subsection{QNN Architecture}

The QNN within our LEP-QNN framework is presented in Fig.~\ref{qnn}, showcasing the methodical arrangement of quantum operations that form the core of our computational model. The circuit is initiated through angle encoding, which effectively transposes each feature of the input data $x_i$ into quantum states:
\begin{equation}
\ket{\psi(x_i)} = \bigotimes_{j=1}^{N} RY(x_{ij} \pi)\ket{0},
\end{equation}
where $RY(\theta)$ is the rotation gate about the Y-axis by an angle proportional to the feature value, which embeds the classical data into a quantum state amenable to quantum processing.

Following the angle encoding, the QNN utilizes an ansatz predetermined yet flexible structure of quantum gates—each parameterized by a set of variables $\theta$.
These gates are responsible for manipulating the qubits to model complex nonlinear relationships within the data:
\begin{equation}
U(\theta)\ket{\psi(x_i)} = \left(\prod_{d=1}^{D} U_d(\theta_d)\right)\ket{\psi(x_i)},
\end{equation}
where $U_d(\theta_d)$ corresponds to the unitary operation at each depth $d$, where $d=5$ to  outline the model's multilayered complexity, with $\theta_d$ representing the set of parameters at that specific layer.  
In designing the ansatz, we review several configurations inspired by existing literature. We implement a pre-testing phase—an algorithmic evaluation that systematically examines different ansatz candidates. In this phase, the following steps are performed:
\begin{itemize}
    \item \textbf{Configuration Sampling:} A set of ansatz candidates is generated, each differing in gate composition and layering. These candidates include various combinations of rotation gates, controlled operations, and entangling layers, designed to capture non-linearities in a diverse manner.
    \item \textbf{Algorithmic Evaluation:} Each candidate is subjected to an initial training cycle using a portion of the loan eligibility dataset. We use performance metrics such as prediction accuracy, convergence speed, and loss reduction as benchmarks. Cross-validation and iterative testing are employed to ensure that the evaluation is robust and that performance remains consistent across multiple data splits.
    \item \textbf{Systematic Comparison:} The performance of each ansatz candidate is compared using a scoring metric that combines accuracy in predicting loan eligibility with the model's capability to represent the essential features of the data.
\end{itemize}
Through this systematic comparison, we identify the fixed ansatz structure as the most effective option for our framework. This choice is based on its superior performance in achieving higher prediction accuracy and its alignment with our methodological objectives—namely, representing the essential features of the data pertinent to loan eligibility. Moreover, the chosen ansatz offers flexibility and can be refined further.
\subsection{Optimizer Analysis}
In refining our LEP-QNN framework, the optimization process is pivotal, serving to calibrate the model's parameters ($\theta$) that minimizes the cost function $C(\theta)$, formulated as:
\begin{equation}
C(\theta) = \frac{1}{N}\sum_{i=1}^{N}(y_i - \hat{y}_i(\theta))^2,
\end{equation}
where $\hat{y}_i(\theta)$ denotes the expected output from the QNN for input instance $i$, and $y_i$ is the corresponding true label.

While gradient descent forms the foundation for this optimization, we explore a suite of advanced optimizers with unique strategies and benefits to enhance the model's learning efficacy and predictive performance. We expand the discussion on the four optimizers utilized within our framework:

\begin{itemize}[leftmargin=*]

\item \textbf{Gradient descent} is used to optimize where the model parameters are iteratively adjusted in the opposite direction of the cost function's gradient:

\begin{equation}
\theta_{t+1} = \theta_{t} - \eta \nabla_\theta C(\theta_{t}),
\end{equation}

where $\eta$ symbolizes the learning rate, dictating the step size toward the minimum of the cost function $C(\theta)$, with $\nabla_\theta C(\theta_{t})$ denoting the gradient at iteration $t$. Despite its simplicity, gradient descent's effectiveness is often limited by its uniform application of the learning rate across all parameters, potentially leading to slow convergence in complex landscapes.

\item \textbf{Adam optimizer} introduces an adaptive learning rate for each parameter, significantly enhancing the convergence rate. It combines the benefits of two other extensions of stochastic gradient descent, AdaGrad, and RMSProp:
\begin{equation}
\theta_{t+1} = \theta_{t} - \frac{\eta}{\sqrt{\hat{v}{t}} + \epsilon} \hat{m}{t},
\end{equation}
where $\hat{m}{t}$ and $\hat{v}{t}$ are estimates of the first (mean) and the second (uncentered variance) moments of the gradients, respectively. Adam automatically adjusts the learning rate for each parameter based on these moments, with $\epsilon$ providing numerical stability. This approach is particularly effective in high-dimensional parameter spaces.

\item \textbf{RMSProp} modifies the learning rate for each parameter based on the moving average of the squared gradients, aiming to scale the gradient inversely with its recent magnitude:
\begin{equation}
\theta_{t+1} = \theta_{t} - \frac{\eta}{\sqrt{E[g^2]{t}} + \epsilon} \nabla C(\theta{t}),
\end{equation}
where $E[g^2]_{t}$ represents the expectation of the squared gradients, stabilizing the optimization in very steep or flat regions of the cost function. RMSProp is especially suitable at navigating complex and varied terrains in loss landscapes. 

\item \textbf{Adagrad} allows the learning rate to adapt based on the parameters, making it larger for infrequent and smaller for frequent parameters:
\begin{equation}
\theta_{t+1} = \theta_{t} - \frac{\eta}{\sqrt{G_{t} + \epsilon}} \nabla C(\theta_{t}),
\end{equation}
where $G_{t}$ is a diagonal matrix where each diagonal element is the sum of the squares of the gradients for each parameter up to iteration $t$. This feature makes Adagrad particularly suitable for dealing with sparse data, as it applies larger updates to less frequent features.
\end{itemize}
These optimizers collectively form the backbone of our optimization strategy for the LEP-QNN framework, each tested for its efficacy in navigating the parameter space toward optimal cost function values. The adaptability of Adam, the gradient scaling of RMSProp, and the parameter-specific learning rates of Adagrad complement the foundational gradient descent approach, offering a robust and comprehensive optimization framework for enhancing the predictive capabilities of our LEP-QNN.

\subsection{Dropout Design}
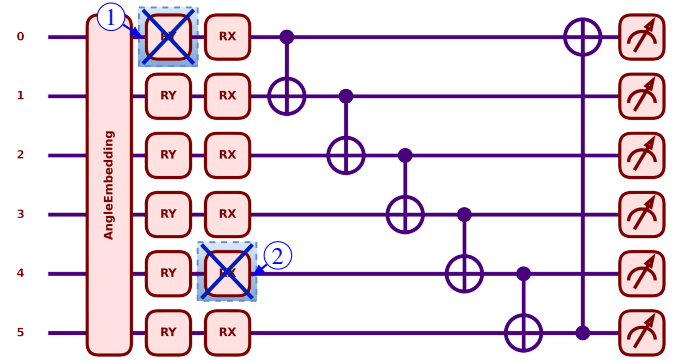
\begin{figure}[htpb]
    \centering
                \begin{tikzpicture}
            \node[anchor=south west, inner sep=0] (image) at (0,0) {
    \includegraphics[width=1\linewidth]{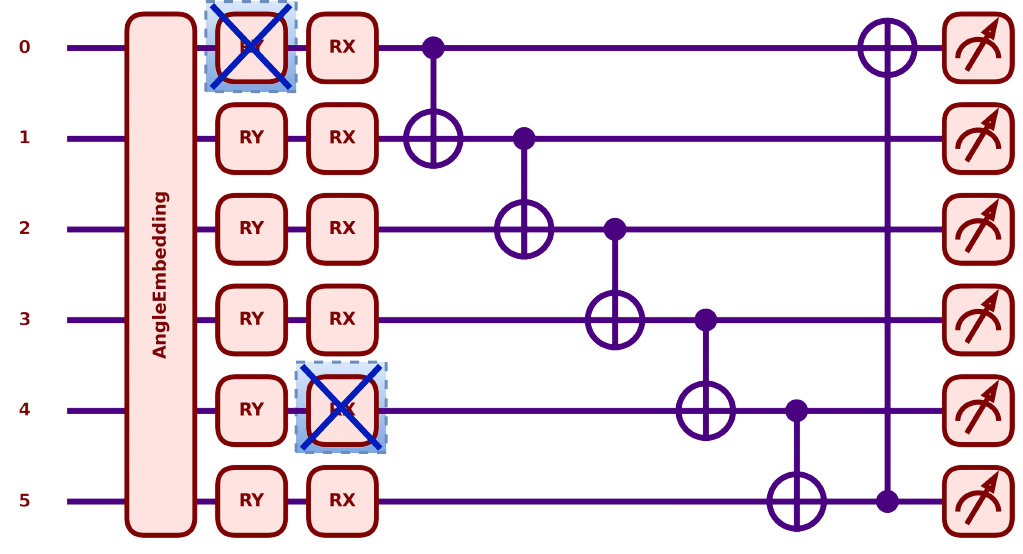}
    };
            \begin{scope}[x={(image.south east)},y={(image.north west)}]
                \node[circle, draw=blue, fill=white, inner sep=1pt] (pointerB) at (0.41,0.31) {\textcolor{blue}{2}};
                \draw[-latex, blue, thick] (pointerB) -- +(-0.04, -0.06); 
                \node[circle, draw=blue, fill=white, inner sep=1pt] (pointerB) at (0.16,0.97) {\textcolor{blue}{1}};
                \draw[-latex, blue, thick] (pointerB) -- +(0.049, -0.059);                 
            \end{scope}
        \end{tikzpicture}
    \caption{QNN circuit implementing dropout, with $RY$ and $RX$ gates randomly deactivated at various layers to prevent overfitting. Gates subject to dropout are marked with blue crosses (\textcircled{\raisebox{-0.9pt}{1}}, and \textcircled{\raisebox{-0.9pt}{2}}), demonstrating the model's reduced sensitivity to specific data features through regularization.}
    \label{dropout}
\end{figure}

A dropout strategy is employed during training to regularize the model and prevent overfitting \cite{srivastava2014dropout}. Dropout is implemented by randomly setting a fraction of the qubits to a fixed state \cite{scala2023general}, thereby reducing the model's sensitivity to particular features of the data:
\begin{equation}
\theta^{'}_{d} =
\begin{cases}
0, & \text{if dropout is applied at depth } d \\
\theta_{d}, & \text{otherwise}
\end{cases},
\end{equation}
where $\theta^{'}_{d}$ is the potentially modified parameter set after applying dropout at depth $d$.
In Fig.~\ref{dropout}, we illustrate the QNN circuit after applying the dropout technique, where certain rotation gates are deactivated, as marked by \textcircled{\raisebox{-0.9pt}{1}} and \textcircled{\raisebox{-0.9pt}{2}}, to enhance the model's ability to generalize and mitigate overfitting. 

The operational flow of our LEP-QNN encompasses a structured procedure from data normalization and state preparation to QNN training and final performance evaluation, as we explained above. The dropout is systematically integrated into the circuit during the training iterations, as Algorithm \ref{alg:lepqnn1} iterates over the QNN parameters, optimizing them by calculating gradients and the subsequent updating mechanism.

In summary, the LEP-QNN framework introduces a highly sophisticated QML approach that combines theoretical quantum mechanics and practical optimization techniques to address the complex problem of loan eligibility prediction. Through rigorous experimentation and careful tuning of the model's parameters, the LEP-QNN stands as a testament to the potential and applicability of quantum-enhanced ML in the financial sector.

\begin{algorithm}[h!]
\label{alg:lepqnn1}
\DontPrintSemicolon
\caption{LEP-QNN}
\KwIn{Training dataset $X_{\text{train}}$ with instances $x_i^{\text{train}}$, training labels $y_{\text{train}}$, testing dataset $X_{\text{test}}$ with instances $x_i^{\text{test}}$, testing labels $y_{\text{test}}$, number of qubits $N$, QNN depth $D$, learning rate $\eta$, number of training iterations $T$.}
\KwOut{MSE Loss $\mathcal{L}$ and performance metrics $\mathcal{M}$ on the testing dataset.}

\textbf{Procedure:}\;
Normalize $X_{\text{train}}$ and $X_{\text{test}}$ to $X_{\text{norm}}^{\text{train}}$ and $X_{\text{norm}}^{\text{test}}$, respectively, using the mean and standard deviation of $X_{\text{train}}$.\;
\For{each instance $x_i$ in $X_{\text{norm}}^{\text{train}}$ and $X_{\text{norm}}^{\text{test}}$}{
    Apply angle encoding to $x_i$ to prepare the quantum state $\ket{\psi(x_i)}$.\;
}
Initialize a set of optimizers $\mathcal{O} = \{\text{GradientDescent}, \text{Adam}, \text{RMSProp}, \text{Adagrad}\}$ with corresponding step sizes.\;
\For{each optimizer $o \in \mathcal{O}$}{
    Randomly initialize QNN parameters $\theta$ for a network of depth $D$.\;
    Initialize an empty list $L_{\text{train}}$ to store the training loss history.\;
    \For{$t = 1$ to $T$}
    {
        Calculate the cost function $C(\theta)$.\;
        Update $\theta$ and calculate loss using $o$.step\_and\_cost with $C(\theta)$, obtaining new parameters $\theta'$ and current loss $l_t$.\;
        Set $\theta = \theta'$ and append $l_t$ to $L_{\text{train}}$.\;
    }
    Compute predictions $P_{\text{test}}$ for $X_{\text{norm}}^{\text{test}}$ using the optimized $\theta^{*}$ without applying dropout, by evaluating $qnn(\ket{\psi(x_i^{\text{test}})}, \theta^{*})$ for each $x_i^{\text{test}}$.\;
    Calculate MSE loss $\mathcal{L}$ and performance metrics $\mathcal{M}$ (e.g., accuracy, precision, recall) for $P_{\text{test}}$ against $y_{\text{test}}$.\;
    Store $\mathcal{L}$ and $\mathcal{M}$ for each optimizer in $\mathcal{O}$ for comparison.\;
}
Identify and report the optimizer yielding the lowest $\mathcal{L}$ and highest performance metrics $\mathcal{M}$ on the testing dataset.\;
\Return{$\mathcal{L}$ and $\mathcal{M}$ for the best-performing optimizer.}\;

\end{algorithm}

\section{Results and Discussion\label{sec4}}
\subsection{Experimental Setup}
In this study, we employ the dataset from Dream Housing Finance Company \footnote{https://www.kaggle.com/datasets/vikasukani/loan-eligible-dataset}, renowned for its home loan services in various regions. The dataset, comprising  614 unique entries for training and 367 for testing, aims to automate loan eligibility through real-time data collected during the customers' application phase. It encompasses parameters such as \textit{Gender, Marital Status, Education, Dependents, Income details, Loan Amount, Loan Term, Credit History, Property Area}, and \textit{Loan Status}, each identified by a \textit{Loan\_ID}. The dataset's breadth enables a comprehensive analysis of the socioeconomic factors influencing loan eligibility, providing an ideal testing environment for our LEP-QNN framework.
 \begin{figure}[htpb]
     \centering
    \includegraphics[width=1\linewidth]{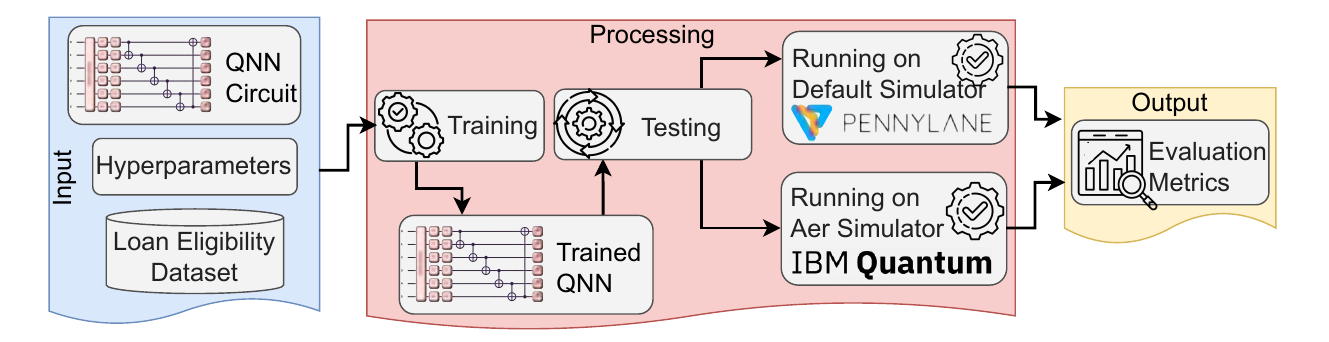}
    \caption{The experimental setup in our evaluation.}
    \label{set}
\end{figure}
Our experiments are conducted using a high-performance computing environment equipped with an Intel(R) Xeon(R) Platinum 8259CL CPU @ 2.50GHz, complemented by access to QC cloud services via the respective software tools, ensuring the rigorous and robust evaluation of our LEP-QNN framework (see Fig. \ref{set}).

In order to optimize our LEP-QNN framework, a variety of hyperparameters and settings are meticulously selected, as detailed in Table \ref{table:hyperparameters}.
We configure the model with six qubits, reflecting the trade-off between computational demand and representational capacity. For visualization purposes, the depth of the QNN is set to 1, while a depth of 5 is employed for model training to capture deeper feature interactions.

\begin{table}[htpb]
\centering
\caption{Hyperparameters and Settings for LEP-QNN.}
\begin{adjustbox}{max width=\linewidth}
\begin{tabular}{p{4cm}|p{4cm}}
\toprule
\textbf{Hyperparameter} & \textbf{Value/Setting} \\ 
\midrule
Number of qubits ($N$) & 6 \\ 
Depth of QNN ($D$) & 1 for visualization, 5 for training \\
Learning rate ($\eta$) & 0.00001 to 0.1 \\
Optimizer($\eta$) & Gradient Descent (0.1), \textbf{Adam} (0.01), RMSProp (0.001), and Adagrad (0.01) \\
Number of iterations ($T$) & 100 \\
Parameters & 60, initialized randomly\\
Train Set Size &  614 instances \\
Test Set Size & 367 instances \\
Dropout rate & Starts at 0.2, decreases by 0.02 each layer \\
Performance Metric & Binary classification metrics \\
Loss Metric & MSE \\
Software Framework & PennyLane and Qiskit \\
\bottomrule
\end{tabular}
\end{adjustbox}
\label{table:hyperparameters}
\end{table}

The learning rate ($\eta$) is carefully chosen within a range of 0.00001 to 0.1 to balance the convergence speed and stability of the training process. The optimizers evaluated include Gradient Descent, Adam, RMSProp, and Adagrad, with the iterative process limited to 100 iterations to maintain computational tractability while ensuring sufficient model refinement. Model parameters are randomly initialized to prevent any bias in the learning process.
To counter potential overfitting, a dropout rate that commences at 0.2 and incrementally decreases by 0.02 with each layer is implemented. The primary performance metrics, centered on binary classification, with MSE as the chosen loss metric, round out our comprehensive approach to evaluating our LEP-QNN's predictive capabilities.

\subsection{Analysis of Optimizers}
The efficacy of our LEP-QNN framework is empirically validated by employing different optimization techniques to estimate the model's predictive prowess. As summarized in Table \ref{tab}, the Adam optimizer emerges as the superior performer, consistently achieving the highest scores across all evaluated metrics. It achieves a notable precision of 0.97, denoting a 97\% proportion of true positive predictions. Furthermore, a recall of 0.98 reflects the model's capacity to correctly identify 98\% of all pertinent cases. The F1-Score stands at 0.97, which underscores the balance between precision and recall. Remarkably, Adam's accuracy reaches 98\%, reinforcing its superiority in accurately forecasting loan eligibility.
RMSProp follows closely, exhibiting a strong performance with precision and recall scores of 0.95 and 0.93, respectively, and an F1-Score of 0.94, leading to an overall accuracy of 96\%. Though RMSProp's results are robust, they are modestly eclipsed by those of the Adam optimizer.
Both Gradient Descent and Adagrad demonstrate consistent outcomes, as indicated by their precision, recall, and F1 scores of 0.94 and 0.93, respectively, with an overall accuracy of 94\%. These findings imply that while these optimizers perform effectively, they do not quite match the benchmark set by Adam and RMSProp in this context.
\begin{table}[htbp]
    \centering
    \caption{Performance metrics of our LEP-QNN.}
    \begin{adjustbox}{max width=\linewidth}
    \begin{tabular}{lcccc}
        \toprule
        \multirow{2}{*}{\textbf{Optimizer}} & \multicolumn{4}{c}{\textbf{Metrics}} \\ 
        \cmidrule(lr){2-5}
        & \textbf{Precision} & \textbf{Recall} & \textbf{F1-Score} & \textbf{Accuracy (\%)}\\
        \midrule
        Gradient Descent & 0.94 & 0.94 & 0.94 & 94 \\
        \midrule
        \textbf{Adam} & \textbf{0.97} & \textbf{0.98} & \textbf{0.97} & \textbf{98}  \\
        \midrule
        RMSProp & 0.95 & 0.93 & 0.94 &  96  \\
        \midrule
        Adagrad & 0.94 &0.93 & 0.93 &  94 \\
        \bottomrule
    \end{tabular}
    \end{adjustbox}
    \label{tab}
\end{table}

\begin{figure}[t]
    \centering
                    \begin{tikzpicture}
            \node[anchor=south west, inner sep=0] (image) at (0,0) {
    \includegraphics[width=1\linewidth]{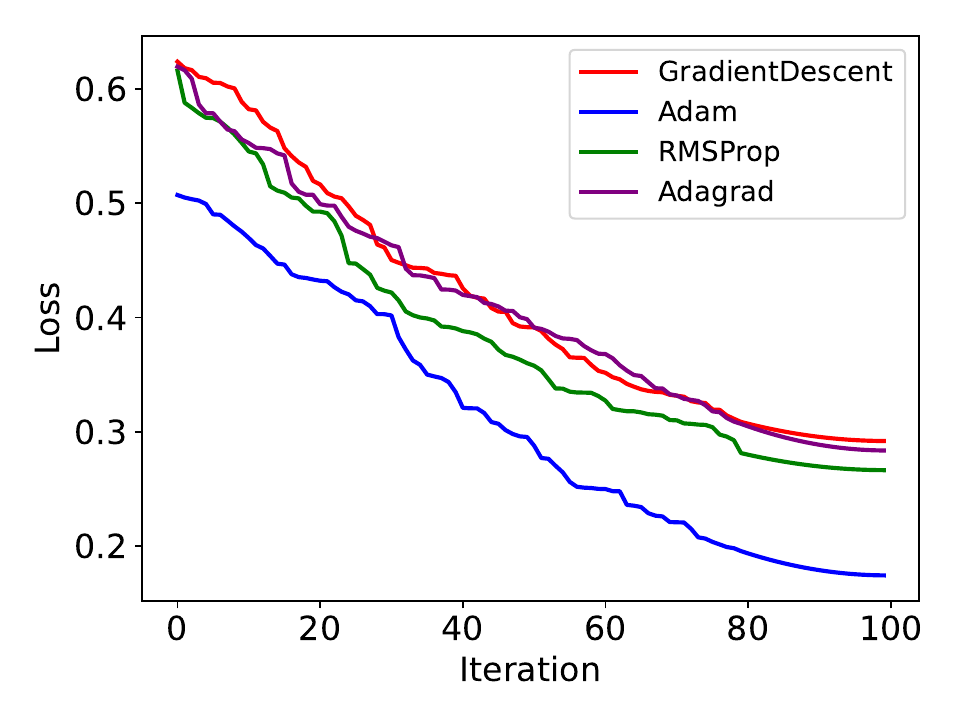}    };
            \begin{scope}[x={(image.south east)},y={(image.north west)}]
                \node[circle, draw=black, fill=white, inner sep=1pt] (pointerB) at (0.77,0.5) {\textcolor{black}{2}};
                \draw[-latex, black, thick] (pointerB) -- +(-0.04, -0.06); 

                \node[circle, draw=black, fill=white, inner sep=1pt] (pointerB) at (0.83,0.295) {\textcolor{black}{1}};
                \draw[-latex, black, thick] (pointerB) -- +(-0.04, -0.07); 
            \end{scope}
        \end{tikzpicture}
        
    \caption{Convergence of loss function for different optimizers. This graph illustrates the loss minimization trajectories over 100 iterations for four optimization algorithms applied within the LEP-QNN Framework. Adam demonstrates the most rapid and stable convergence towards a lower loss, suggesting a more efficient optimization process than the other techniques. The described trends indicate the effectiveness of each optimizer in refining the QNN model for loan eligibility prediction.}
    \label{loss}
\end{figure}

The convergence behavior of the loss function over consecutive training iterations directly indicates the LEP-QNN framework's learning efficiency. Fig.~\ref{loss} shows the distinctive convergence patterns for each optimizer within our framework.
As shown, the Adam optimizer exhibits a pronounced and steadfast decline in loss, realizing the most acute gradient among the optimizers. Adam achieves swift loss reduction from the onset, as marked by \textcircled{\raisebox{-0.9pt}{1}}, maintaining this trend throughout the optimization phase. This rapid convergence signifies Adam's efficacious parameter space exploration, refining the QNN model swiftly and precisely.
In contrast, gradient descent demonstrates a steady yet relatively moderate reduction in loss. This tempered pace implies that a more significant number of iterations may be requisite for gradient descent to achieve the optimization levels attained by Adam, potentially necessitating more computational time and resources, particularly when applied to larger datasets or more intricate models.
RMSProp and Adagrad delineate an intermediate trajectory, with RMSProp closely following Adam's performance, notably in the latter iterations, as indicated by \textcircled{\raisebox{-0.9pt}{2}}, and Adagrad displays a marginally better loss reduction rate than gradient descent. The behaviors of RMSProp and Adagrad reinforce their robustness as alternatives, although not quite rivalling Adam's optimization efficiency.

\subsection{Analysis of Noise Models}
\begin{figure}[t]
    \centering
                    \begin{tikzpicture}
            \node[anchor=south west, inner sep=0] (image) at (0,0) {
    \includegraphics[width=1\linewidth]{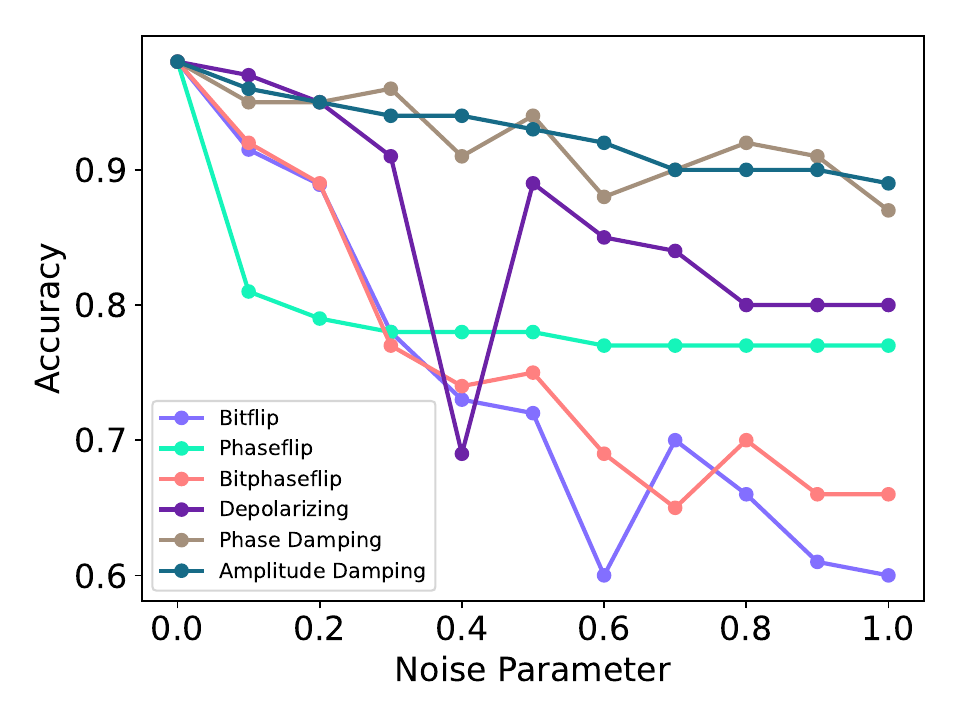}    };
            \begin{scope}[x={(image.south east)},y={(image.north west)}]

                \node[circle, draw=blue, fill=white, inner sep=1pt] (pointerB) at (0.29,0.66) {\textcolor{blue}{1}};
                \draw[-latex, blue, thick] (pointerB) -- +(-0.0, 0.07); 
                \coordinate (startPoint) at (0.21,0.77);
                \draw[draw=blue] (startPoint) arc[start angle=-120, end angle=-80, radius=2cm];
                \node[circle, draw=blue, fill=white, inner sep=1pt] (pointerB) at (0.75,0.25) {\textcolor{blue}{2}};
                \draw[-latex, blue, thick] (pointerB) -- +(0.08, 0.00004); 
                \draw[-latex, blue, thick] (pointerB) -- +(0.08, 0.09);

                \node[circle, draw=blue, fill=white, inner sep=1pt] (pointerB) at (0.9,0.64) {\textcolor{blue}{3}};
                \draw[-latex, blue, thick] (pointerB) -- +(-0.0, 0.067);
                \draw[-latex, blue, thick] (pointerB) -- +(0.01, -0.062); 
                \coordinate (startPoint) at (0.88,0.716);
                \draw[draw=blue] (startPoint) arc[start angle=-120, end angle=-80, radius=1cm];
            \end{scope}
        \end{tikzpicture}
    
    \caption{Accuracy of the LEP-QNN under various noise models. This graph describes the effect of six quantum noise models on the accuracy of the LEP-QNN framework. Each curve represents the model's accuracy against increasing noise parameters, demonstrating the varying degrees of the robustness of the LEP-QNN to bitflip, phaseflip, bitphaseflip, depolarizing, phase damping, and amplitude damping noise. The model's resilience to different noise intensities is critical for its application in quantum-enhanced predictive analytics.}
    \label{noise}
\end{figure}

The resilience of QNNs to quantum noise is important for their practical application, as noise can significantly undermine performance. Fig.~\ref{noise} explains the influence of diverse noise models on the accuracy of the LEP-QNN framework relative to noise intensity.
The graph exhibits that our LEP-QNN framework sustains high accuracy at low noise levels, with all noise types yielding above 90\% accuracy for noise parameters under 0.2, as shown in \textcircled{\raisebox{-0.9pt}{1}}. Upon escalation of the noise parameter, a decline in accuracy is observed universally across noise models. Bitflip and bitphaseflip, in particular, demonstrate a stark decrement in accuracy as the noise parameter approaches unity as marked by \textcircled{\raisebox{-0.9pt}{2}}, highlighting the model's heightened sensitivity to these noise forms.
Conversely, depolarizing, phase damping, and amplitude damping models present a more tempered decline in accuracy with augmenting noise, as indicated by \textcircled{\raisebox{-0.9pt}{3}}, with the latter two maintaining superior accuracy over an extensive range of noise parameters, denoting intrinsic resilience within our LEP-QNN framework to these noise types.
\subsection{Comparison with Existing Works}
The comparison of existing approaches, as detailed in Table \ref{table:comparison}, highlights the groundbreaking accuracy achieved by our LEP-QNN framework. With an accuracy of 98\%, our framework significantly outperforms conventional models using the same dataset, marking a substantial advance in predictive analytics for loan eligibility. The nearest competitor, an ensemble method comprising the best three classical algorithms, lags by 7.74\%—a clear indication of the quantum model's superior capability in managing complex data. This comparison attests to the novelty of our approach, given that prior works have not explored QML models for this dataset and loan prediction in general. The superior performance of LEP-QNN paves the way for future explorations into quantum models in the financial sector and beyond. 
\begin{table}[htpb]
\caption{Comparison of existing works on the same dataset.}
\centering
\begin{adjustbox}{max width=\linewidth}
\begin{tabular}{lll}
\toprule
\textbf{Reference}&\textbf{Approaches} & \textbf{Accuracy (\%)} \\
\midrule
\cite{uddin2023ensemble}&Ensemble with Best Three & 87.26 \\
\cite{anand2022prediction}&Extra Trees & 86.2 \\
\cite{kumar2022customer}&Decision Tree with AdaBoost & 84 \\
\cite{blessie2019exploring}&Naïve Bayesian & 80.4 \\
\textbf{Our approach}&\textbf{LEP-QNN} & \textbf{98} \\
\bottomrule
\end{tabular}
\end{adjustbox}
\label{table:comparison}
\end{table}
\subsection{Discussion}
In the discussion, we examine the implications of these findings more profoundly. The resilience of our LEP-QNN framework to quantum noise is not uniform; certain noise types have a more deleterious effect on model performance. For instance, bitflip and bitphaseflip noise types may necessitate the development of specialized noise mitigation strategies to preserve the integrity of predictions. The relative robustness to depolarizing, phase damping, and amplitude damping noises could indicate important characteristics of our LEP-QNN framework that could be further exploited to improve its noise resistance.
The precipitous drop in accuracy observed at a noise parameter of 0.4 in the depolarizing noise model invites further investigation into the threshold effects and potential stabilization of model performance in noisy environments. Such insights will be critical for developing QML models that can operate reliably in the presence of quantum noise, particularly in Noisy Intermediate-Scale Quantum (NISQ) devices.
The performance of various optimizers also has far-reaching implications for the practical deployment of QNNs. Our results support the Adam optimizer's prioritization in similar QML tasks due to its swift convergence and high accuracy, contributing to a more efficient and effective loan eligibility prediction process. These findings are instrumental in advancing the field of QML, paving the way for robust and reliable quantum-enhanced predictive analytics.
\section{Conclusion\label{sec5}}
Concluding this study on QML applied to financial predictive analytics, our LEP-QNN framework demonstrates the significant impact of QC in enhancing the accuracy of financial models. With an impressive 98\% accuracy rate in loan eligibility predictions, LEP-QNN highlights the advanced capabilities of QML in processing complex financial data and establishes new benchmarks for applying quantum technologies in data analysis.
At the core of our work is the development of a specialized QNN architecture. The inclusion of a unique dropout mechanism within this architecture is a critical advancement in overcoming overfitting, enhancing the model's generalization across diverse financial scenarios. This feature underscores the practical value of quantum-enhanced models in real-world applications.
Our comprehensive evaluation of different optimizers has shed light on optimal strategies to enhance QNN performance. This aspect of our research guided us to achieve optimum model efficiency and highlighted the delicate balance between quantum algorithms and performance enhancement. Moreover, examining the model's endurance against various quantum noise models offers insights into its stability and reliability within the unpredictable nature of QC environments.
As we wrap up, our LEP-QNN framework marks a milestone for financial analytics, showcasing QC as the key to innovation and precision in the field. This work opens doors to future explorations in QML, inviting further research into quantum technologies' potential to address complex analytical challenges across different sectors. Through this study, we lay the groundwork for future advancements in QC applications, encouraging ongoing exploration of quantum technologies in analytics and beyond.
\section*{Acknowledgments}
This work was supported in part by the NYUAD Center for Quantum and Topological Systems (CQTS), funded by Tamkeen under the NYUAD Research Institute grant CG008.
\bibliographystyle{IEEEtran}

\bibliography{IEEE.bib}

\end{document}